\documentclass[article,preprint,floatfix]{revtex4-1}
\usepackage{textcomp}
\usepackage{graphicx}
\usepackage{amssymb}
\usepackage{epsfig}
\usepackage{ifsym}
\usepackage{wasysym}
\usepackage{latexsym}

\begin{document}

\title{Unforeseen high temperature and humidity stability of FeCl$_3$ intercalated few layer graphene}

\author{Dominique Joseph Wehenkel$^\dag$}
\author{Thomas Hardisty Bointon$^\dag$}
\author{Tim Booth$^\ddag$}
\author{Peter B{\o}ggild$^\ddag$}
\author{Monica Felicia Craciun$^\dag$}
\author{Saverio Russo$^\dag$}
\email{Correspondence to S.Russo@exeter.ac.uk}
\affiliation{$^\dag$ Centre for Graphene Science, College of Engineering, Mathematics and Physical Sciences, University of Exeter, Exeter EX4 4QF, UK\\
$^\ddag$ Center for Nanostructured Graphene (CNG), Department of Micro- and Nanotechnology Engineering, Technical University of Denmark, DK-2800 Kongens Lyngby, Denmark}

\begin{abstract}
We present the first systematic study of the stability of the structure and electrical properties of FeCl$_3$ intercalated few-layer graphene to high levels of humidity and high temperature. Complementary experimental techniques such as electrical transport, high resolution transmission electron microscopy and Raman spectroscopy conclusively demonstrate the unforseen stabiliy of this transparent conductor to a relative humidity up to $100 \%$ at room temperature for 25 days, to a temperature up to $150\,^\circ$C in atmosphere and up to a temperature as high as $620\,^\circ$C in vacuum, that is more than twice higher than the temperature at which the intercalation is conducted. The stability of FeCl$_3$ intercalated few-layer graphene together with its unique values of low square resistance and high optical transparency, makes this material an attractive transparent conductor in future flexible electronic applications.  
\end{abstract}

\maketitle

\textbf{\large{Introduction}}
\\
Transparent conductive electrodes are extensively used for optoelectronic applications such as solar cells \cite{gunes2007,ameri2009}, light emitting devices (LED) and displays \cite{lewis2004,tak2002,Wu1997}. To date, indium tin oxide (ITO) is the industrial standard for transparent electrodes with a high transmittance (T) of $85 \% $ in the visible wavelength range  and a square resistance (R$_{sq}$) as low as $ 10 \Omega / \rm{sq}  $ for a film thickness of $10 \, \mu \rm{m}$ \cite{Lee2008, De2008}. However this material has a number of limiting properties which are holding back the developement of conceptually new types of screen technologies. For example, ITO strongly absorbs the infrared radiation, making it an interesting material largely for applications in the visible range of the light spectrum. Furthermore, ITO is relatively brittle \cite{Lee2008, De2008} accounting for the rigidity of nowadays screens. Hence new materials are needed to let the screen technology emerge from the confines of walls and become foldable and wearable. 

The chemical functionalization of graphene with atoms or molecules provides a valuable way to engineer the desired combination of physical properties needed to enable the aforementioned novel technologies \cite{bae2010,Bao2014,Khrapach2012,craciun2013}. For example, FeCl$_3$ intercalated few-layer graphene has a square resistance lower than ITO (just $\approx\,8\,\Omega /\rm{sq}$) with an optical transmittance of $92\,\%$ measured in bilayers from ultra-violet up to the near infra-red wavelength range \cite{Khrapach2012}.  This functionalization is easily scalable to large area substrates using graphene obtained with different methods such as epitaxial growth on silicon carbide (4H-SiC)\cite{bointon2014}.  The unique combination of low square resistance, high optical transmittance together with the exceptional mechanical flexibility of graphene makes this hybrid material a valuable alternative substitute of ITO as a transparent electrode. However, it is well known that FeCl$_3$, a commonly used etchant for copper, is highly hygroscopic and soluble in water. Intuition might suggest that FeCl$_3$ functionalized graphene would not be stable in air, with the humidity present in air rapidly diminishing the integrity and properties of the material. 

Here we demonstrate that FeCl$_3$ intercalated few-layer graphene (FeCl$_3$-FLG) is an unexpected highly stable form of functionalized graphene. This conclusion is based on a systematic comparative study of the evolution of the Raman spectra, high resolution transmission electron microscopy (HRTEM), and electrical transport characteristics upon exposing FeCl$_3$-FLG to high relative humidity ($\rm{H}\,>90\,\%$ and up to $100\,\%$) at room temperature, to a temperature up to $150\,^\circ$C in atmosphere and up to a temperature as high as $620\,^\circ$C in vacuum for HRTEM. After three weeks continual exposure to a relative humidity ranging from 95 to 100 $\%$ at room temperature, we observe that the square resistance and the Raman spectra are unchanged. HRTEM measurements up to a temperature of 620$^\circ$C in vacuum, that is more than twice higher than the temperature at which intercalation is conducted, show no measurable change in the structure of the material. The stability of FeCl$_3$-FLG to high humidity and temperatures widens considerably the range of potential applications targetted by graphene materials. 

\textbf{\large{Results}}
\\
Few layer graphene are prepared by micromechanical cleavage of natural graphite on standard Si/SiO$_2$ substrates and transmission electron microscopy (TEM) grids. The functionalization with FeCl$_3$ is conducted using an established vapour transport method in a two-zone furnace \citep{Dresselhaus2002,Khrapach2012}. The stage of intercalation represented by the number of carbon layers separating two subsequent intercalated layers, is determined using Raman spectroscopy \cite{Khrapach2012}. The electrical properties are characterized using multi-terminal devices fabricated with standard eletron-beam lithography, deposition of Cr/Au ($10/50\,\rm{nm}$) followed by lift-off, see inset in Figure 1a. The electrical resistance is measured (1) \textit{in situ} while controlling the humidity in a closed chamber and (2) in air after heating the samples using a hotplate.

\begin{figure*}
\centering
\includegraphics[scale=0.7]{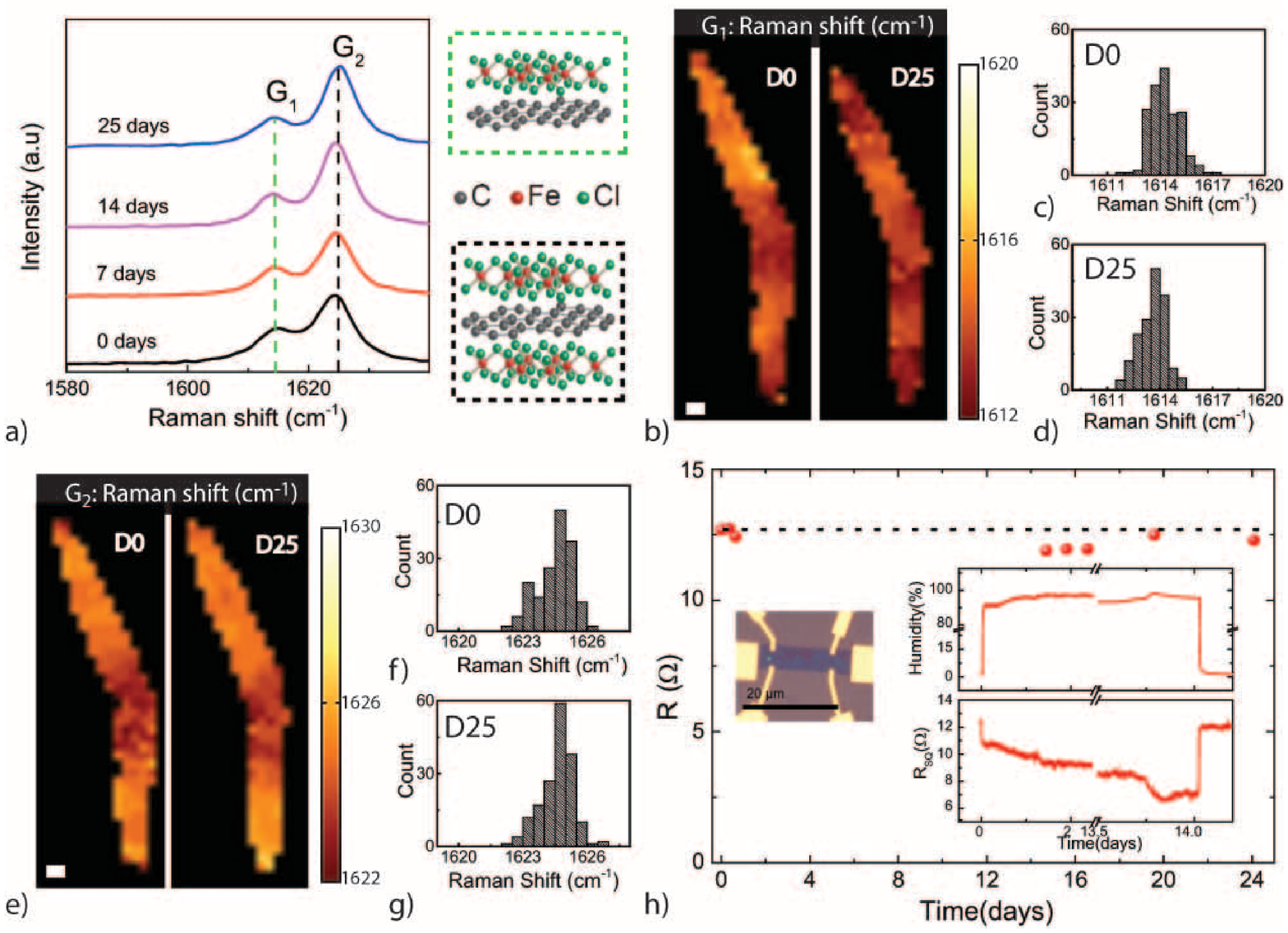}
\caption{}
\end{figure*}

Figure 1a shows the G-peaks of the Raman spectrum at the same location of a representative FeCl$_3$-FLG after exposing the sample to an atmosphere with relative humidity $\rm{H}\,>\,95\%$ for various days as indicated in the graph. In contrast to the case of pristine graphene for which a single G-peak \cite{Ferrari2006}  is measured at $1585\,\rm{cm}^{-1}$ (G$_0$), in the intercalated material this peak shifts to 1615 $\rm{cm}^{-1}$ (G$_1$) and 1625 $\rm{cm}^{-1}$ (G$_2$) can be observed. Both G$_1$ and G$_2$ are the consequence of stiffening of the E$_{2\rm{g}}$ phonon mode \cite{Zhan2010,Zhao2011} caused by charge transfer from FeCl$_3$ to graphene. More specifically, G$_1$ is characteristic of a graphene layer being doped by only one adjacent FeCl$_3$ layer whereas G$_2$ corresponds to a graphene layer sandwiched between two layers of FeCl$_3$ (see crystal structure illustrations in Figure 1a). No measurable shift of the $G_1$- and $G_2$-peaks  is observed even after exposing the flake for 25 days to high humidity, suggesting that the intercalated compound is indeed not affected by the humidity.

To demonstrate that the structure of FeCl$_3$-FLG is stable against prolonged exposure to extremly high levels of humidity, we conduct a detailed study of the Raman maps of the $G_1$- and $G_2$-peaks before and after exposing a representative flake to $\rm{H}>95\%$ for 25 days, see Figure 1b-g. We find that overall the position of the aforementioned Raman peaks does not change significantly. Indeed, the distribution of the Raman shifts of $G_1$ (see Figure 1c-d) and $G_2$ (see Figure 1f-g) before and after exposure to humidity only exhibit at most a shift of $\approx 1 \rm{cm}^{-1}$ that is within the accuracy of the Raman spectroscopy tool used for this experiment.

\begin{figure}[ht]{}
\includegraphics[width=0.7\textwidth]{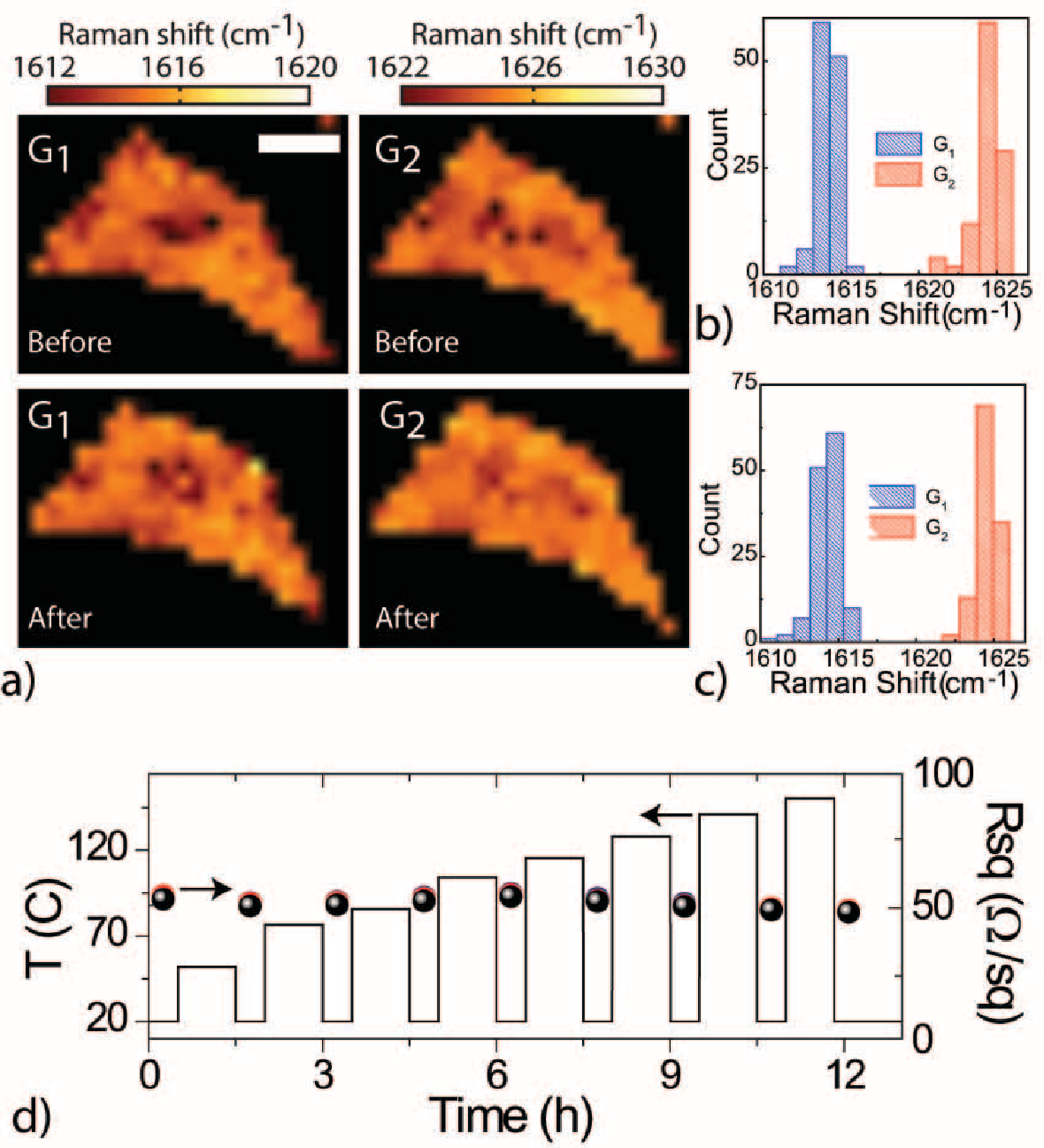}
\caption{}
\end{figure}

Having established that the structure of this intercalated compound does not change upon exposure to humid atmosphere, we proceed to characterize the stablity of the electrical properties under the same experimental conditions. Figure 1h shows a plot of the measured R$_{\rm{sq}}$ in dry atmosphere after subsequent exposures to high levels of humidity for the sample shown in the micrograph image in the inset. We observe that the initial value of R$_{\rm{sq}} \approx 13\,\Omega/\rm{sq}$ is unchanged after exposing the sample to $\rm{H}>95\%$ for 25 days. Furthermore, \textit{in situ} measurements of R$_{\rm{sq}}$ while exposing the device to high levels of humidity show that R$_{\rm{sq}}$ decreases when the sample is exposed to high levels of humidity (see graph in the inset of Figure 1h). This drop in resistance is a reversible process, since the initial value of R$_{\rm{sq}}$ is restored in the sample in dry atmosphere. These observations suggest that water molecules condensed on the surface of FeCl$_3$-FLG might contribute to an increase of electrical conductivity without causing irreversible changes to the material. Insight in the microscopic origin behind the stability of this material in atmosphere can be gained when considering that in bulk intercalated graphite the carbon interlayer spacing of non intercalated regions in close proximity to intercalated regions have values similar to the pristine case \cite{Thomas1980}. Consequently, the large interlayer binding energy characterizing these non-intercalated regions can effectively act as a diffusion barrier for intercalants. In FeCl$_3$-FLG a similar mechanism could occur near the edges of the flakes, whereby a narrow de-intercalated edge would block the diffusion of FeCl$_3$ molecules out of the structure.

To further evaluate the suitability of FeCl$_3$-FLG for future electronic applications, we also need to characterize the stability of this material to high temperatures. Also in this case we conduct a comparative study of Raman spectroscopy and electrical transport characterization before and after heating FeCl$_3$-FLGs in atmosphere on a hot plate. Figure 2a shows the colour coded maps of the Raman shift of G$_1$ and G$_2$ before (top graphs) and after (bottom graphs) heating the sample for 1h on a hotplate in atmosphere at $100\,^{\circ}\mathrm{C}$. It is apparent that Raman shifts of just a few $\rm{cm}^{-1}$ are measured, corresponding to the accuracy of the spectrometer. This is more clearly seen when comparing the corresponding histograms of the Raman shifts for G$_1$ and G$_2$ peaks before (Figure 2b) and after heating the sample (Figure 2c). The electrical transport measurements also show no significant change of the room temperature square resistance after heating the sample to subsequently higher temperatures from $50\,^{\circ}\mathrm{C}$ up to $150\,^{\circ}\mathrm{C}$ in multiple steps of 1h duration, see Figure 2d.

\begin{figure}[ht]{}
\includegraphics[width=0.7\textwidth]{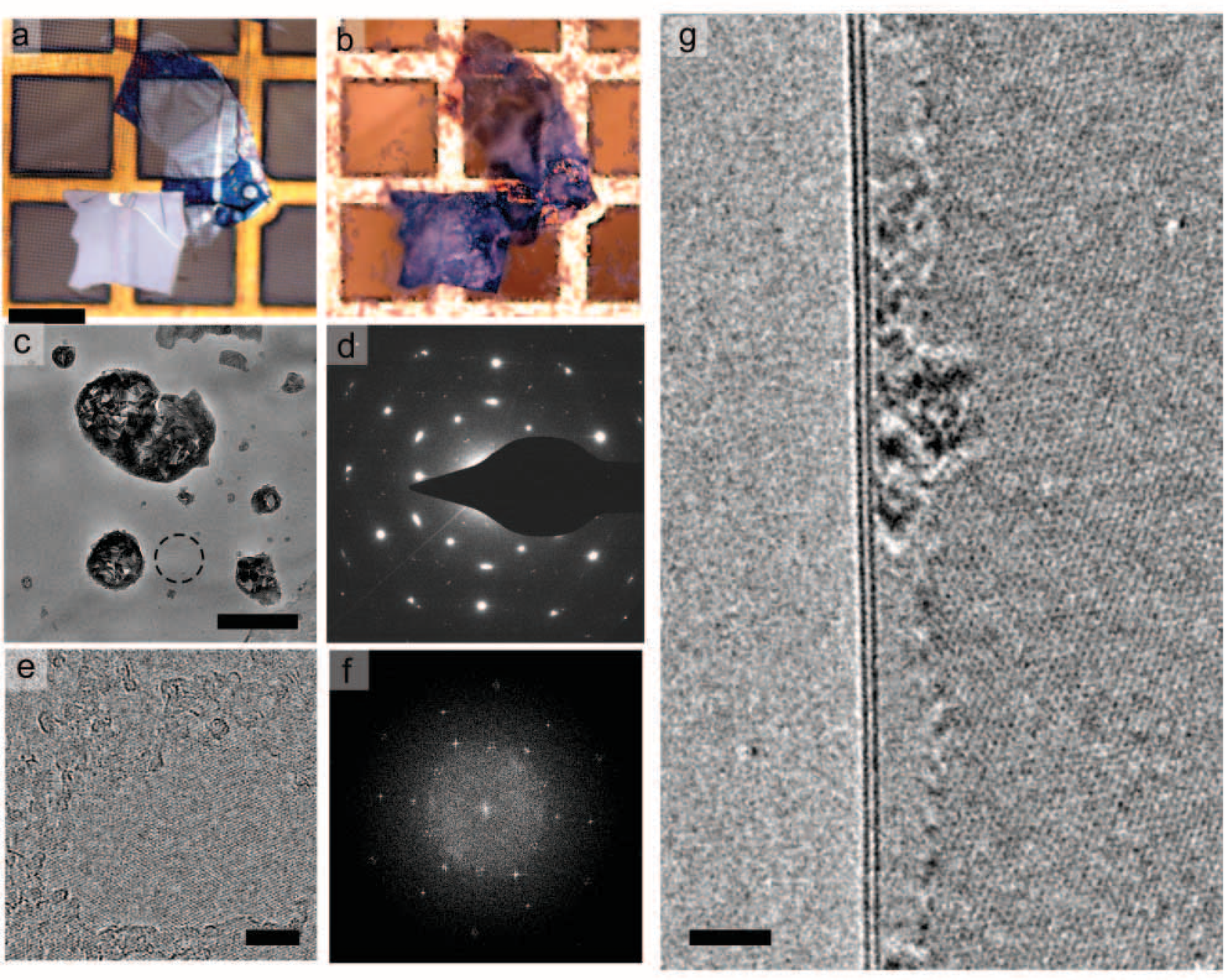}
\caption{}
\end{figure}

\textbf{\large{Discussion}}
\\
To elucidate the microscopic origin of the thermal stability we have conducted a study of the structure of FeCl$_3$-FLG upon heating up to $620\,^{\circ}\mathrm{C}$ in vacuum with an FEI Titan E-Cell 80-300ST aberration-corrected transmission electron microscope equipped with an inconel-based heating holder. A few layer graphene flake was transferred to a standard TEM grid (see Fig. 3a) using a published technique \cite{Meyer2008} and intercalated with FeCl$_3$ with the two zone method (see Fig. 3b-c). Imaging is performed at 80 keV to reduce the effects of knock-on damage \cite{Meyer2012}. After intercalation additional lattice periodicities are visible in the multilayer flake and can be observed both in selected area diffraction patterns and high-resolution imaging (Fig. 3d-f). These periodicities are also visible in a folded bilayer region of the intercalated graphene flake (Fig. 3g), proving that this is a stage I intercalated compound. More specifically, the intercalated material is visible throughout the bilayer region except for some 1-2 nm away from the edge of a folded bilayer, which is not intercalated and that is measured \textit{prior} to any heating process. It is likely that a small radius of curvature in the bilayer graphene results in the exclusion of intercalated material here - the width of this observed exclusion region corresponds well to the typically observed minimum diameters of double-walled carbon nanotubes \cite{Allen2012}. 

The TEM studies also show on the surface of the suspended graphene flake the presence of nanoparticles, probably consisting of FeCl$_x$, of less than 250 nm diameter, resulting from an excess of intercalant species condensing on the surface of graphene (Fig. 3c). During heating of the sample up to $620\,^{\circ}\mathrm{C}$ over 1500 sec at a pressure of 10$^{-5}$-10$^{-6}$ mbar and concurrent imaging no change is observed in the structure of the sample - FeCl$_3$ intercalated bilayer graphene retains additional periodicities due to the graphene and intercalant (Supplementary Video S1). We note that the nanoparticles visible on the surface of intercalated graphene also do not change structure during this ramped heating. FeCl$_3$ has a melting point of $315\,^{\circ}\mathrm{C}$, as compared to the melting point of FeCl$_2$ of $677\,^{\circ}\mathrm{C}$, indicating that the particles on the surface are likely formed by FeCl$_2$. These studies show that the intercalation of FeCl$_3$ is stable at least up to $620\,^{\circ}\mathrm{C}$ in vacuum.

In conclusion, we demonstrate that FeCl$_3$ intercalated few-layer graphene is highly stable to high levels of humidity and to high temperature. This is supported by a systematic comparative study of the measurements obtained from three complementary techniques: (1) Raman spectroscopy, (2) high resolution transmission electron microscopy, and (3) electrical transport. The Raman spectra and the square resistance of this material are unchanged upon exposing continually FeCl$_3$-FLG to an atmosphere with relative humidity up to 100$\%$ at room temperature for at least 25 days, and to a temperature up to $150\,^\circ$C in atmosphere. At the same time high resolution transmission microscopy confirms that the structure of the material is unaffected by heating FeCl$_3$-FLG up to $620\,^{\circ}\mathrm{C}$ in vacuum.  The surprising stability of the structure and electrical properties of FeCl$_3$-FLG together with its unique values of low square resistance and high optical transparency, makes this material an attractive replacement for ITO in future transparent and flexible electronic applications.

\textbf{\large{Methods}}\\
\textbf{Measurement techniques.} The Raman spectra where measured in air and at room temperature with a Renishaw spectrometer using a 532 nm laser wavelength with a 1.5 $\mu$m spot size and 1 mW of incident power. \\
The electrical measurements where conducted in a four terminal geometry using an AC current bias ($1\mu$A and 11Hz frequency) while the voltage was recorded using phase sensitive lock-in detection. \\
To detect the relative humidity a  HIH-4000 Humidity sensor from Honeywell. The relative humidity was controlled by passing dry nitrogen or ambient air through DI water in a gas bubbler. 

\textbf{\large{Acknowledgements}}\\ 
We acknowledge D. Horsell for letting us use a vacuum microscope stage, F. Withers for helping in the preparation of samples for HRTEM and A. De Sanctis for useful discussions on the interpretation of Raman spectroscopy measurements. SR and MFC acknowledge financial support from EPSRC (Grant no. EP/J000396/1, EP/K017160/1, EP/K010050/1, EP/G036101/1, EP/M001024/1, EPM002438/1).

\textbf{\large{Author contributions}}\\ 
MFC and SR conceived and directed the experiment. DJW conducted the fabrication, Raman and electrical measurements. THB produced intercalated few-layer graphene and conducted the electrical measurements at high temperature in atmosphere. TB conducted the HRTEM measurements. TB and PB interpreted the HRTEM measurements.  

\textbf{\large{Additional information}}\\
\textbf{Supplementary Information} accompanies this paper at http://www.nature.com/scientificreports.\\
\textbf{Competing financial interests:} the authors declare no competing financial interests.
License: This work is licensed under a Creative Commons
Attribution-NonCommercial-NoDerivs 3.0 Unported License. To view a copy of this
license, visit http://creativecommons.org/licenses/by-nc-nd/3.0/

\newpage
\textbf{\large{Figure legends}} 

\textbf{Figure 1.}a) Shows a plot of Raman spectra measured in a rapresentative FeCl$_3$-FLG sample before (0 days) and after exposure to $\rm{H}\,>\,95\%$ for 7, 14 and 25 days shifted for clarity along the y-axis. The peaks G$_1$ and G$_2$ are highlighted on the graph, and the corresponding crystal structure is shown in the illustrations on the right side of the graph. Panel (b)  shows the colour coded Raman maps of the G$_1$ peak before (D0) and after (D25) exposure to $\rm{H}\,>\,95\%$. The white scale bar corresponds to $2\,\mu \rm{m}$. (c) and (d) are the corresponding hystograms of the Raman shift of G$_1$ for D0 and D25 respectively. Graphs in (e) are colour coded Raman maps of G$_2$ before (D0) and after (D25) exposure to high levels of humidity. The white scale bar corresponds to $2\,\mu \rm{m}$. (f) and (g) are graphs of the histograms of the Raman shift of G$_2$ for D0 and D25 respectively. The main graph in (h) is a plot of the value of R$_{\rm{sq}}$ in dry atmosphere after exposing the device shown in the micrograph picture in the inset to high levels of humidity for different intervals of time. The graphs in the inset show the \textit{in situ} R$_{\rm{sq}}$ \textit {vs.} time and the relative levels of humidity.

\textbf{Figure 2.} a) Shows four colour coded Raman maps of the G$_1$ and G$_2$ peaks before (top panels) and after (bottom panels) heating of a representative FeCl$_3$-FLG to $100\,^{\circ}\mathrm{C}$ for 1h. The white scale bar corresponds to $5\,\mu \rm{m}$. The graphs in (b) and (c) are the corresponding histograms of the Raman shift of G$_1$ and G$_2$ before and after heating the sample. The plots in (d) show the values of R$_{\rm{sq}}$ measured at room temperature (top graph) after heating the sample to subsequently higher temperatures for 1h (bottom graph). The black, red and blue data points refer to three different choices of contact probes on the same flake.

\textbf{Figure 3.} a),b) Optical images of few-layer graphene flake before and after intercalation. Scale bar 100 $\mu \rm{m}$. c) Low magnification image of intercalated multilayer graphene. Nanoparticles of FeCl$_x$ can be seen on the surface. Scale bar 250 nm. d) Selected area diffraction pattern of region indicated in (c). e) High resolution image of FeCl$_3$-intercalated few layer graphene after heating to 850 K. Scale bar 5 nm. f) Fourier transform of region in (e). g) Edge of folded bilayer after heating to 850 K - FeCl$_3$ periodicity is visible up to 1-2 nm from the edge of the folded bilayer. Scale bar 2 nm.


\begin{thebibliography}{4}
\bibitem{gunes2007} 
G\"{u}nes, S., Neugeauer, H. \& Sariciftci, N. S. Conjugated polymer-based solar cells. \textit{Chem. Rev.} \textbf{107}, 1324-1338 (2007).

\bibitem{ameri2009} 
Ameri, T., Dennler, G., Lungenschmied, C. \& Brabec, C. J. Organic tandem solar cells: A review.  \textit{Energy Environ. Sci.} \textbf{2}, 347-363 (2009). 

\bibitem{lewis2004} 
Lewis, J., Grego, S., Chalamala, B., Vick, E. \& Temple, D. Highly flexible transparent electrodes for organic light-emitting diode-based displays.  \textit{App. Phys. Lett.} \textbf{85}, 3450-3452 (2004).

\bibitem{tak2002} 
Tak, Y. H., Kim, K. B., Park, H. G., Lee, K. H. \& Lee, J. R. Criteria for ITO (indium–tin-oxide) thin film as the bottom electrode of an organic light emitting diode. \textit{Thin Solid Films} \textbf{411}, 12-16 (2002).

\bibitem{Wu1997} 
Wu, C. C., Wu, C. I., Sturm, J. C. \& Kahn, A. Surface modification of indium tin oxide by plasma treatment: An effective method to improve the efficiency, brightness, and reliability of organic light emitting devices. \textit{App. Phys. Lett.} \textbf{70},  1348-1350 (1997).

\bibitem{Lee2008} 
Lee, J. Y., Connor, S. T., Cui, Y. \& Peumans, P. Solution-processed metal nanowire mesh transparent electrodes. \textit{Nano Lett.} \textbf{8},  689-692 (2008).

\bibitem{De2008}  
De, S. \& Coleman, J. N. Are there fundamental limitations on the sheet resistance and transmittance of thin graphene films?  \textit{ACS Nano} \textbf{4},  2713-2720 (2010).


\bibitem{cairns2000} 
Cairns, D. R., \textit{et al.} Strain-dependent electrical resistance of tin-doped indium oxide on polymer substrates. \textit{App. Phys. Lett.}\textbf{76},  1425-1427 (2000).

\bibitem{yu2011} 
Yu, Z., \textit{et al.} Highly Flexible Silver Nanowire Electrodes for Shape-Memory Polymer Light-Emitting Diodes. \textit{Adv. Matter.} \textbf{23}, 664-668 (2011).

\bibitem{bae2010} 
Bae, S., \textit{et al.}  Roll-to-roll production of 30-inch graphene films for transparent electrodes. \text{Nat. Nanotechnol.} \textbf{5}, 574-578 (2010).

\bibitem{Bao2014} 
Bao, W., \textit{et al.} Approaching the limits of transparency and conductivity in graphitic materials through lithium intercalation. \textit{Nat. Comm.} \textbf{5},  4224 (2014).

\bibitem{Khrapach2012} 
Khrapach, I., \textit{et al.}  Novel highly conductive and transparent graphene-based conductors.\textit{Adv. Mater.} \textbf{24},  2844-2849 (2012).

\bibitem{craciun2013}
Craciun, M. F., Khrapach, I., Barnes, M. D. \& Russo, S. Properties and applications of chemically functionalized graphene. \textit{J. Phys. Condens. Matter.} \textbf{25},  423201 (2013).
    
\bibitem{bointon2014} 
Bointon, T. H., \textit{et al.} Approaching Magnetic Ordering in Graphene Materials by FeCl$_{3}$ Intercalation \textit{Nano Lett.} \textbf{14},  1751-1755 (2014).

\bibitem{kong2014}
Song, Y., Fang, W., Hsu, A. L. \& Kong, J. Iron (III) Chloride doping of CVD graphene. \textit{Nanotechnology} \textbf{25}, 395701 (2014).


\bibitem{Dresselhaus2002} 
Dresselhaus, M. S. \& Dresselhaus, G. Intercalation compounds of graphite. \textit{Adv. Phys.} \textbf{51} , 1-186 (2002).


\bibitem{Ferrari2006} 
Ferrari, A. C., \textit{et al.} Raman spectrum of graphene and graphene layers. \textit{Phys. Rev. Lett.} \textbf{97},  187401 (2006).


\bibitem{Zhan2010} 
Zhan, D., \textit{et al.} FeCl$_3$-based few-layer graphene intercalation compounds: single linear dispersion electronic band structure and strong charge transfer doping. \textit{Adv. Func. Matter.} \textbf{20},  3504-3509 (2010).


\bibitem{Zhao2011} 
Zhao, W., Tan, P. H., Liu, J. \& Ferrari, A. C. Intercalation of few-layer graphite flakes with FeCl$_3$: Raman determination of fermi level, layer by layer decoupling, and stability. \textit{J. Am. Chem. Soc.} \textbf{133},  5941-5946 (2011).

\bibitem{Thomas1980}
Thomas , J. M., Millward, G. R., Schlogl, R. F. \& Boehm, H. P. Direct imaging of a graphite intercalate: evidence of interpenetration of 'stages' in graphite: ferric chloride. \textit{Mat. Res. Bull.}, \textbf{15}, 671-676 (1980).

\bibitem{Meyer2008} 
Meyer, J. C., Girit, C. O., Crommie, M. F. \& Zettl A. Hydrocarbon lithography on graphene membranes. \textit{Appl. Phys. Lett.} \textbf{92}, 123110 (2008).

\bibitem{Meyer2012} 
Meyer, J. C., \textit{et al.} Accurate measurement of electron beam induced displacement cross sections for single-layer graphene. \textit{Phys. Rev. Lett.} \textbf{108},  196102 (2012).

\bibitem{Allen2012}  
Allen, C. S., Robertson, A. W. , Kirkland, A. I. \& Warner, J. H. The identification of inner tube defects in double-wall carbon nanotubes. \textit{Small} \textbf{8},  3810-3815 (2012).

\end{thebibliography}
\end{document}